\documentclass[11pt, a4paper]{article}

\usepackage[utf8]{inputenc}
\usepackage[T1]{fontenc}
\usepackage{authblk} 
\usepackage{amsmath, amssymb, amsfonts}
\usepackage{graphicx}
\usepackage{hyperref}
\usepackage{geometry}
\usepackage{bm}
\usepackage{algorithmic}
\usepackage{algorithm}
\usepackage{booktabs}
\usepackage{multirow}
\usepackage{caption}
\usepackage{nomencl}
\makenomenclature

\newcommand{\copyrightnotice}{%
 \thanks{This is a Gold Open Access article made available under the CC-BY license. Published in \textit{Expert Systems with Applications}. DOI: \href{https://doi.org/10.1016/j.eswa.2025.129145}{10.1016/j.eswa.2025.129145}}
}

\title{\textbf{Multiagent Reinforcement Learning in Enhancing Resilience of Microgrids under Extreme Weather Events}\copyrightnotice}

\author[1]{Yin Wu}
\author[2]{Wei-Yu Chiu\thanks{Corresponding author: weiyu.chiu@deakin.edu.au}}
\author[1]{Yuan-Po Tsai}
\author[3]{Shangyuan Liu}
\author[3]{Weiqi Hua}

\affil[1]{Department of Electrical Engineering, National Tsing Hua University, Taiwan}
\affil[2]{School of Information Technology, Deakin University, Australia}
\affil[3]{School of Engineering, University of Birmingham, B15 2TT, U.K.}

\date{} 

\begin{document}

\maketitle

\begin{abstract}
Grid resilience is crucial in light of power interruptions caused by increasingly frequent extreme weather events. Well-designed energy management systems (EMS) have made progress in improving microgrid resilience through the coordination of distributed energy resources (DERs), but still face significant challenges in addressing the uncertainty of load demand caused by extreme weather. The integration of deep reinforcement learning (DRL) into EMS design enables optimized microgrid control strategies for coordinating DERs. Building on this, we proposed a cooperative multi-agent deep reinforcement learning (MADRL)-based EMS framework to provide flexible scalability for microgrids, enhance resilience and reduce operational costs during power outages. Specifically, the gated recurrent unit with a gating mechanism was introduced to extract features from temporal data, which enables the EMS to coordinate DERs more efficiently. Next, the proposed MADRL method incorporating action masking techniques was evaluated in the IEEE 33-Bus system using real-world data on renewable generation and power load. Finally, the numerical results demonstrated the superiority of the proposed method in reducing operating costs as well as the effectiveness in enhancing microgrid resilience during power interruptions.
\end{abstract}

\vspace{1em}

\newpage
\section{Introduction}\label{sec_Intro}
\subsection{Background and Motivation} 
In recent years, rapid global climate change has intensified extreme weather events, leading to a higher frequency of natural disasters that disrupt electrical infrastructure \cite{01Teresa}, \cite{01Jibonananda}, \cite{01MF}. Among many disasters, typhoons merit particular attention for their recurrent and severe impact on power grids, particularly in coastal and islanded regions. The high winds accompanying typhoons often result in extensive damage to crucial components of microgrids, such as feeder lines and transformers, leading to widespread power outages \cite{01Zhang}. For instance, in 2014, typhoon "Rammasun" caused extensive damage to the power grid in Hainan, China, leading to widespread power outages and the collapse of 27 transmission towers, with estimated economic losses of over 10 billion CNY \cite{01AnLi}. In August 2017, Hurricane "Harvey" caused massive power outages in Texas, USA, affecting approximately 300,000 residents who endured prolonged blackouts lasting for several weeks \cite{01Kavousi}. Similarly, in 2023, Typhoon "Koinu" brought heavy rainfall to Taiwan, resulting in 190 injuries and leaving over 510,000 households without electricity. Therefore, the sensitivity of electrical infrastructure to extreme typhoon events has highlighted the necessity for resilient power grid systems to mitigate their impact \cite{01Emanuele}.

Concurrently, renewable energy adoption and energy storage systems (ESS) as distributed energy resources (DERs) have gradually increased to cope with environmental changes, thereby providing significant benefits to power grids \cite{10akorede}. However, due to the decarbonisation of the power sector, the rising penetration of renewable energy sources and power loads poses challenges to the power system's stability, scheduling and management \cite{19AlamEMreview}. To solve these problems, it is crucial to improve the restoration capabilities of grids during typhoon weather \cite{16Wang}, \cite{22younesi}. The potential operational measures to enhance power network resilience are diverse, encompassing preventive islanding \cite{Hayez2022}, microgrids \cite{Hussain2019}, storage \cite{Dugan2021}, topology reconfiguration \cite{Ding2017ARM}, and others \cite{05Noebels}. Basically, introducing microgrids is considered to provide a key solution for enhancing grid resilience \cite{10Govind}.

Microgrid technology can function in islanded mode, reduce dependence on the main grid and improve power system robustness by decentralizing features \cite{LEVORATO2022117471}, \cite{CHAKRABORTY20154643}, which is particularly beneficial during typhoon-induced power outages. Key objectives for resilient microgrids include maximizing load survivability and minimizing energy costs during islanded operation \cite{20wang}. Therefore, the energy management system (EMS) with dynamic control is essential for achieving these objectives \cite{18zia}.

Given the urgent need to enhance the resilience of microgrids under extreme weather conditions and the limitations in adaptability of traditional optimization-based strategies, this study proposes an EMS framework that integrates gated recurrent units (GRUs) and multi-agent deep reinforcement learning (MADRL) to address these shortcomings. The main contributions are summarized as follows:

\begin{enumerate}
\item This research is the first to integrate GRUs into a DRL framework dedicated to microgrid energy management. In contrast to static models typically employed in previous studies, our approach significantly improves the ability of EMS to predict and adapt to fluctuations in renewable energy generation and load demand by extracting temporal information utilising GRUs. It enables more reliable energy allocation, in particular under unpredictable conditions.

\item This study distinctively considers the impacts of typhoons and other extreme weather events. This approach allows for a more realistic and flexible response strategy that enhances the ability of microgrids to maintain operations under uncertain and changing weather conditions.

\item This study introduces the concept of multiple breakpoints in microgrid topology, which has not been extensively explored in previous research. These breakpoints represent the potential points of failure or disconnection in the power grid network. By accounting for the probabilities of multiple breakpoints, the proposed strategy offers a more granular and realistic assessment of grid resilience. Compared to the commonly used MINLP and DDPG algorithms, the proposed method reduces load shedding per day by around 2.4 MW and 1.22 MW respectively.
\end{enumerate}

The rest of the article is organized as follows.
Section~\ref{sec_model} identifies mathematical models for the operation of DERs, EMS as well as GRU algorithms.
In section~\ref{sec_method}, a MARL-based energy management method for improving microgrid resilience is proposed.
In Section~\ref{sec_result}, comprehensive numerical experiments are conducted to verify the superiority of the method in comparison with existing methods.
Finally, this article is concluded in section~\ref{sec_conclu}.

\subsection{Literature Review} 
In general, there are two methods to improve microgrid resilience through EMS, which are rule-based and optimization-based strategies respectively \cite{09Xie}. Rule-based EMS manages microgrids using predefined criteria or logic rules. In contrast, optimization-based strategies use mathematical optimisation methods to provide better performance for microgrid operation by minimising an objective function of operating costs including fuel, startup/shutdown, curtailment and emission costs. \cite{Restrepo2021}
Although rule-based strategies are easily implemented, they may not lead to the optimal solutions. Optimization-based strategies rely on analytical or numerical algorithms to find optimal solutions in complex and dynamic environments. \cite{10Roy}.
A multi-stage resilient scheduling model under tropical cyclones is introduced, employing a dual dynamic integer programming method to ensure robustness against unforeseen outage uncertainties \cite{18Qiu}.
Gan \textit{et al.} \cite{18gan} demonstrated an energy-sharing scheme that connected multiple microgrids, preventing an ESS from over-discharging and reducing the system uncertainties. Younesi \textit{et al.}\cite{21younesi} developed a two-stage scheduling strategy: the first stage was an economic operation for the next day, and the second stage was a real-time operation maintaining the resilience in natural disasters, reflecting the economics aspect in the resilience-oriented objective.
Linear programming was used in \cite{19Poudel}, \cite{18tavakoli}, \cite{14Khodaei} to utilize distributed power generation sources to supplement critical loads during disasters while not violating the power grid’s connectivity constraints.
Zhao \textit{et al.} \cite{15Zhao} proposed the distributed coordinated reinforcement learning (DCRL) algorithm, which integrates distributed consensus control with reinforcement learning to dynamically adjust distributed controller gains for optimizing the voltage control performance of microgrids. This becomes particularly crucial in addressing load fluctuations triggered by extreme weather conditions within power systems.
In \cite{16Gholami} and \cite{20wu}, stochastic programming was further applied to both grid-connected and islanded scenarios; both methods addressed uncertain parameters, but required using complete mathematical models.

However, conventional optimization methods have two main drawbacks. 
First, knowledge about a complete mathematical system model or system parameters is generally needed.
Second,  the uncertainties in PV generations and power loads can be difficult to address \cite{21chen}.
As such, deep reinforcement learning (DRL) has received much attention in power grid management due to its adaptivity \cite{18Zhang}.
For instance, Zhao \textit{et al.} \cite{22zhao} proposed a learning capacity enhancement method based on a convolutional neural network (CNN) with a multi-buffered double-deep Q-network, which enhances the long-term resilience of the power system.
In \cite{21Kamruzzaman}, a MADRL framework was presented to select the correct size and location of reactive power compensators and mitigate voltage violations by deploying shunt resources during windstorms.  Dehghani \textit{et al.} \cite{21dehghani} devised a DRL-based strategy that replaced wood utility poles in a large-scale power system impacted by hurricanes, assuming that the environment needed to be fully observable. Moreover, a resilient control approach for multi-energy microgrids was proposed by replacing the deterministic network in traditional DRL with a Bayesian probabilistic network to approximate the value function distribution, effectively addressing the Q-value overestimation issue \cite{34Tingqi}.
However, these methods have primarily concentrated on enhancing long-term grid resilience, there is still room for improvement in the research of real-time control and immediate support for power loads under specific weather extremes.

Several studies have examined a reinforcement learning-based EMS. The authors in  \cite{02Huang} introduced a deep Q-network mode-free DRL approach to optimise the microgrid formation strategy, which enables resilient distribution networks through microgrid formation under frequent extreme climate events.
In \cite{21zhou}, a DRL framework was proposed to determine DERs' energy scheduling while allowing disconnection of microgrid components based on typhoon trajectory estimation.

A DRL-based EMS using hurricane parameters analysis was proposed in \cite{21hosseini}, where a sequential decision-making control problem was reformulated to address the scalability of DERs.
Wang \textit{et al.} \cite{22wang} developed a multiagent DRL algorithm to coordinate multiple mobile ESS, which is challenging due to the necessity of considering transportation for ESS.
In addition, power supply and load redistribution services were considered common primary objectives. 
In \cite{21bedoya} and \cite{22du}, a DRL framework was developed to control DERs to support remaining loads and restore service simultaneously after a power outage occurrence.

Although existing MADRL-based EMS frameworks have contributed to microgrid management, these methods still have limitations when addressing real-time energy management under extreme weather conditions, which are unable to effectively tackle the rapid fluctuations in renewable energy generation and load demand. 
In recent years, CNN and recurrent neural networks (RNN) have demonstrated significant advantages in time series modelling, particularly in processing the complex time-series features from PV generation and load prediction data \cite{xiang2024short}.
GRU networks, as an efficient RNN architecture, can selectively capture and retain long-term dependencies in time series data, which is especially suitable for data prediction scenarios with uncertainty.
In addition, compared with RNN models, GRU mitigates the vanishing gradient problem through its unique gating mechanism, thereby improving the learning ability of the model for complex time series data and further enhancing the real-time prediction and decision-making capabilities of the MADRL framework under extreme weather conditions.

A microgrid usually has multiple DERs to be controlled, such as generators and ESS. When several controllable DERs exist, single-agent reinforcement learning encounters a large action space, and the performance can degrade. In this case, multi-agent reinforcement learning (MARL) based algorithms offer significant strengths \cite{01LYU}, MARL maintains high-performance levels in microgrid environments with multiple controllable DERs through the reduction of the action space for individual agents, parallel learning, cooperative strategies, and improving the system robustness \cite{02Dawei}.
Furthermore, agents in MARL can gain benefit from sharing experience with each other \cite{08busoniu}.
An actor-critic algorithm combines the advantages of policy gradient and value-based methods. In general terms, the actor refers to a policy gradient method that can deal with a continuous action space \cite{99konda}, and the critic refers to a value-based method that can improve the learning efficiency by leveraging temporal difference approaches \cite{99sutton}.
In addition, deterministic policy algorithms are considered more data efficient than stochastic policy algorithms \cite{14silver}. This is because deterministic policy simplifies the process of estimating the policy gradient, enabling faster and more stable learning and improvement by eliminating the estimation variance \cite{lopez2023}.
Therefore, multiagent deep deterministic policy gradient (MADDPG) \cite{17lowe} 
provided a suitable framework for multiagent  energy management, which 
uses network outputs as actions in the microgrid, such as charging and discharging control. The policy network outputs need to be in a valid range;
action clipping is a common way of ensuring the feasible range, but it  
impacts the learning process negatively because of the action saturation. 
Another way is to 
map the network outputs into valid ranges (sometimes called action masking) \cite{22Huang}, but a careful design of the mapping is needed.

Motivated by the potential of learning-based methods, 
we propose a MADDPG-based algorithm for energy management to improve microgrid resilience in the presence of typhoons.
Energy storage systems in the electricity network are considered responsible for energy scheduling to maintain a minimal operational cost during power outages incurred by typhoons.
Predictions of PV and load are fed into a GRU model \cite{14cho} to reduce the state dimension and extract temporal information.  
An action masking method dynamically adjusts the ESS charging/discharging range over time, thereby enhancing learning efficiency
The microgrid considers disconnection from the main grid due to typhoons for energy management. The proposed algorithm was compared with single-agent DDPG, mixed-integer nonlinear programming (MINLP), multi-agent proximal policy optimization (MAPPO), and a rule-based method. Numerical results demonstrate that our method outperforms all baseline approaches in terms of operational cost during power outages.

\section{System Models and Problem Formulation}\label{sec_model}
This section presents a model of a microgrid; The model includes PV, ESS, generators, and an EMS. When the microgrid has disconnected from the main grid because of the violent storm, the EMS can operate ESS and generators and shed some loads to maintain resilience.

\subsection{Microgrid Model} 
Natural disasters, such as floods, thunderstorms, earthquakes, and typhoons can bring destructive damage to power grids. 
In such circumstances, a microgrid may disconnect from the main grid. 
When the microgrid is in islanded mode,  it only relies on DERs to meet power demands.
Therefore, the usage of DERs is essential to maintain the resilience of a microgrid \cite{08Huang}. 
In this work, the microgrid is assumed to be equipped with DERs, such as PV, ESS, and diesel generators.

\subsubsection{Photovoltaic Module}
PV power generation depends on weather conditions, which incurs much uncertainty and is often considered as uncontrollable.
The power generations in time slot $t$ are denoted by $P_{i,t}^{\text{PV}}$, where $i$ represents the PV module index.
The constraints of PV modules can be expressed as
\begin{equation}\label{eq PV bound}
0 \leq P_{i,t}^{\text{PV}} \leq P_{i,\text{max}}^{\text{PV}}
\end{equation}
where $P_{i,\text{max}}^{\text{PV}}$ represent the power generation capacity of PV module $i$.

\subsubsection{Energy Storage System}
The microgrid is equipped with some ESS that store surplus PV energy or energy drawn from the main grid. 
Let $SoC_{j,t}$ and $P_{j,t}^{\text{ESS}}$ denote the state of charge and charging/discharging power of the $j$th ESS at time $t$, respectively. The state of charge $SoC_{j,t}$ satisfies
\begin{equation}\label{eq ESS SoC bound}
SoC_{j}^{\text{min}} \leq SoC_{j,t} \leq SoC_{j}^{\text{max}}
\end{equation}
to prevent overcharging or over-discharging.
The constraints of ESS can be expressed as
\begin{equation}\label{eq ESS power bound}
P_{j,\text{min}}^{\text{ESS}} \leq P_{j,t}^{\text{ESS}} \leq P_{j,\text{max}}^{\text{ESS}}
\end{equation}
where  $P_{j,t}^{\text{ESS}}>0$ represents a charging event and $P_{j,t}^{\text{ESS}}<0$ represents a discharging event, 
and $P_{j,\text{max}}^{\text{ESS}}$ and $P_{j,\text{min}}^{\text{ESS}}$ represent charging and discharging power limits, respectively.

The state of charge $SoC_{j,t}$ is affected by charging or discharging energy $P_{j,t}^{\text{ESS}} \Delta t$, where ${\Delta t}$ represents the duration of a time slot. The energy storage dynamic can be expressed as
\begin{equation}\label{eq ESS SoC dynamic}
SoC_{j,t+1} = 
\left\{
  \begin{array}{ll}
SoC_{j,t} + \delta^{\text{dis}}\frac{P_{j,t}^{\text{ESS}} \Delta t}{E_{j}^{\text{ESS}}},  \hbox{ if } \displaystyle P_{j,t}^{\text{ESS}} \leq 0;    \\
SoC_{j,t} + \delta^{\text{ch}}\frac{P_{j,t}^{\text{ESS}} \Delta t}{E_{j}^{\text{ESS}}},  \quad \hbox{ otherwise }
  \end{array}
\right.
\end{equation}
where $E_{j}^{\text{ESS}}$ in MWh represents the energy capacity, and $\delta^{\text{ch}}$ and $\delta^{\text{dis}}$ represent the charging and discharging efficiency of the ESS in the context of the integrated energy system.

\subsubsection{Diesel Generators}
Generators support the microgrid to satisfy power demand in islanded mode.
Assuming that the power factor of the microgrid is close to 1 \cite{Akarne2023}, the reactive power is set to zero and only active power is considered \cite{15comodi}.
The power generation of generator $i$ in time slot $t$ is denoted by $P_{k,t}^{\text{gen}}$, satisfying
\begin{equation}\label{eq generator bound}
P_{k,\text{min}}^{\text{gen}} \leq P_{k,t}^{\text{gen}} \leq P_{k,\text{max}}^{\text{gen}}
\end{equation}
where $P_{k,\text{max}}^{\text{gen}}$ and $P_{k,\text{min}}^{\text{gen}}$ are the upper and lower power limits of the generator, respectively.

\subsubsection{Load}
Power load changes over time.
The power demand of load  $l$ in time slot $t$ is denoted as $P_{l,t}^{\text{load}}$.
Shedding load in an emergency response is a common operation to withstand a power outage, primarily applied to industrial loads. In residential communities, shedding load is generally avoided as it can negatively affect the well-being of residents.
We denote the parameter
\begin{equation}\label{eq load shed st}
\alpha_{l,t} \in [0,1]
\end{equation}
as the proportion of load shedding, where $\alpha_{l,t}=0$ means the load $l$ at a specific time $t$ is completely supplied without shedding.

\subsection{Energy Management System}
An EMS is integrated into the microgrid, which can control several components such as ESS and generators.
The EMS also collects temporal information, such as predictions of PV generation and power load.
During typhoons, the microgrid may disconnect from the main grid. 
The EMS must utilize the knowledge about the possibility of the microgrid disconnecting from the main grid.
Let $\hat{F}_{t}$ denote the probability of microgrid disconnection, satisfying   
\begin{equation}\label{eq fragility dist}
0 \leq \hat{F}_{t} \leq 1.
\end{equation}

Based on this information, the EMS can control generators and ESS more efficiently.
The EMS can utilize power from the main grid to charge ESS and cover the power loads when the microgrid is still connected to the main grid.
On the other hand, the EMS can control generators and discharge ESS to support the power loads or shed some loads if needed.
Regardless, the total power must be in balance to meet physical constraints including voltage and frequency stability, and grid interconnection limits. The power balance equation in time slot $t$ can be expressed as
\begin{equation}\label{eq power balance}
\begin{split}
0=\sum_{l\in \mathcal{L}}(1-\alpha_{l,t}) P_{l,t}^{\text{load}}-\sum_{i\in \mathcal{I}}P_{i,t}^{\text{PV}}+\sum_{j\in \mathcal{J}}P_{j,t}^{\text{ESS}}\\
-\sum_{k\in \mathcal{K}}P_{k,t}^{\text{gen}}-P_{t}^{\text{grid}}
\end{split}
\end{equation}
here $P_{t}^{\text{grid}}>0$ represents power flowing  from the main grid, and $P_{t}^{\text{grid}}<0$  represents power flowing back to the main grid,
and $\mathcal{L}$, $\mathcal{I}$, $\mathcal{J}$, and $\mathcal{K}$ are the index sets of load, PV, ESS, and generators, respectively.

The objectives of an EMS are to meet power loads as effectively as possible and minimize operating costs. Meeting power loads helps microgrids maintain a stable power supply during sudden demand increases or supply interruptions, reducing the impact of outages on customers. Minimizing operating costs allows microgrids to allocate resources efficiently, cutting unnecessary expenses and ensuring economic sustainability. Achieving these goals enhances microgrid resilience against unexpected events. 
Resilience $R$ can be inversely related to the total cost incurred from load shedding. Specifically, we focus on the penalty of load shedding term in the total operational cost equation. To define resilience mathematically:
\begin{equation}\label{eq resilience}
\begin{split}
R \propto-\sum_{l \in L} \alpha_{l, t} \lambda^{\text {lood }} P_{l, t}^{\text {load}}
\end{split}
\end{equation}
The operational cost includes the cost of ESS degradation $\lambda^{\text{ESS}}$,  the cost of generator usage $\lambda^{\text{gen}}$, the transaction with the main grid $\lambda^{\text{grid}}$, and the penalty of load shedding $\lambda^{\text{load}}$.
The operational cost (total cost) in time slot $t$ can be expressed as 

\begin{equation}\label{eq cost cal}
\begin{split}
C_{t}^{\text{total}}=[\sum_{j\in \mathcal{J}}\lambda^{\text{ESS}}\left\vert\mathop{\min}\{P_{j,t}^{\text{ESS}},0\}\right\vert+
\sum_{k\in \mathcal{K}}\lambda^{\text{gen}}P_{k,t}^{\text{gen}}\\
+\lambda^{\text{grid}} \left\vert P_{t}^{\text{grid}}\right\vert+\sum_{l\in \mathcal{L}}\alpha_{l,t}\lambda^{\text{load}}P_{l,t}^{\text{load}}] \Delta t.
\end{split}
\end{equation}

By suitably controlling the DERs, the EMS reduces operational costs during unplanned power outages.
The load shedding cost $\lambda^{\text{load}}$ directly impacts the amount of load shedding. Larger $\lambda^{\text{load}}$ leads to a smaller amount of load being unserved, improving the microgrid resilience.
The EMS can manage the charging/discharging of an ESS $P_{j,t}^{\text{ESS}}$ and the power supply of a generator $P_{k,t}^{\text{gen}}$ to support power load; load shedding can be performed if necessary.
The goal of the EMS is to minimize the operational cost by solving
\begin{equation}\label{eq objective}
\begin{split}
\mathop{\min}_{\substack{P_{j,t}^{\text{ESS}}, P_{k,t}^{\text{gen}}, \alpha_{l,t}\\ j\in \mathcal{J},k\in \mathcal{K},l\in \mathcal{L}}}\sum_{t}C_{t}^{\text{total}}\\
\text{subject to ~(\ref{eq PV bound})--(\ref{eq power balance})}.
\end{split}
\end{equation}
Because of the uncertainty of power outages as well as prediction errors involved, learning-based methods are generally desired. By leveraging techniques like LSTM networks, Bayesian networks, and Q-learning, these approaches enable autonomous decision-making, fault diagnosis, and policy learning without the need for accurate system models \cite{JoshiSurvey2023}.

\subsection{GRU Model}
GRU is a type of RNN architectures that use gating mechanisms to regulate the flow of information to mitigate the issues of gradient exploding or gradient vanishing and enabling better learning in earlier layers. GRU and LSTM are both proficient at handling temporal data, with GRU being favoured for its faster processing and comparable performance with fewer parameters than LSTM \cite{01Olivieri}. In the context of optimizing renewable energy usage by predicting PV and load data, a GRU-based model is deployed. This model utilizes current PV generation, power load, future PV predictions, and future load predictions as inputs at each time slot $t$.
The output is a characteristic vector $\mathbf{v}_{t}$ representing the input data.
The vector is used for energy scheduling of an ESS.
The characteristic vector can be expressed as
\begin{equation}\label{eq GRU}
\mathbf{v}_{t} = GRU(
\left[
  \begin{array}{cccc}
    P_{1,t}^{\text{PV}} & \hat{P}_{1,t+1}^{\text{PV}} & \cdots & \hat{P}_{1,t+T-1}^{\text{PV}} \\
    \vdots & \vdots & \ddots & \vdots \\
    P_{\left\vert\mathcal{I}\right\vert,t}^{\text{PV}} & \hat{P}_{\left\vert\mathcal{I}\right\vert,t+1}^{\text{PV}} & \cdots & \hat{P}_{\left\vert\mathcal{I}\right\vert,t+T-1}^{\text{PV}} \\
    P_{1,t}^{\text{load}} & \hat{P}_{1,t+1}^{\text{load}} & \cdots & \hat{P}_{1,t+T-1}^{\text{load}} \\
    \vdots & \vdots & \ddots & \vdots \\
    P_{\left\vert\mathcal{L}\right\vert,t}^{\text{load}} & \hat{P}_{\left\vert\mathcal{L}\right\vert,t+1}^{\text{load}} & \cdots & \hat{P}_{\left\vert\mathcal{L}\right\vert,t+T-1}^{\text{load}}
  \end{array}
\right])
\end{equation}
where $T$ represents the observation window, and
$\hat{P}_{i,t}^{\text{PV}}$ and $\hat{P}_{l,t}^{\text{load}}$ are the predictions of PV generation and power load, respectively, and
$\mathcal{I}$ and $\mathcal{L}$ represent the index sets of PV systems and load respectively.

\section{Proposed MADRL Methodology}\label{sec_method}
This section develops a deep reinforcement learning framework 
that enhances the resilience of a Microgrid in consideration of 
the uncertainties in PV generations, power load, and power outages.
In this framework, each ESS is considered as an agent that interacts with the environment jointly.
To develop an algorithm for the EMS, we first present the structure of MADRL.
Then, a GRU model is introduced to handle the temporal information, such as predicted PV generation and power load.
Finally, the details of the MADRL algorithm are discussed.

\begin{figure}
\centering
  \includegraphics[width=8cm]{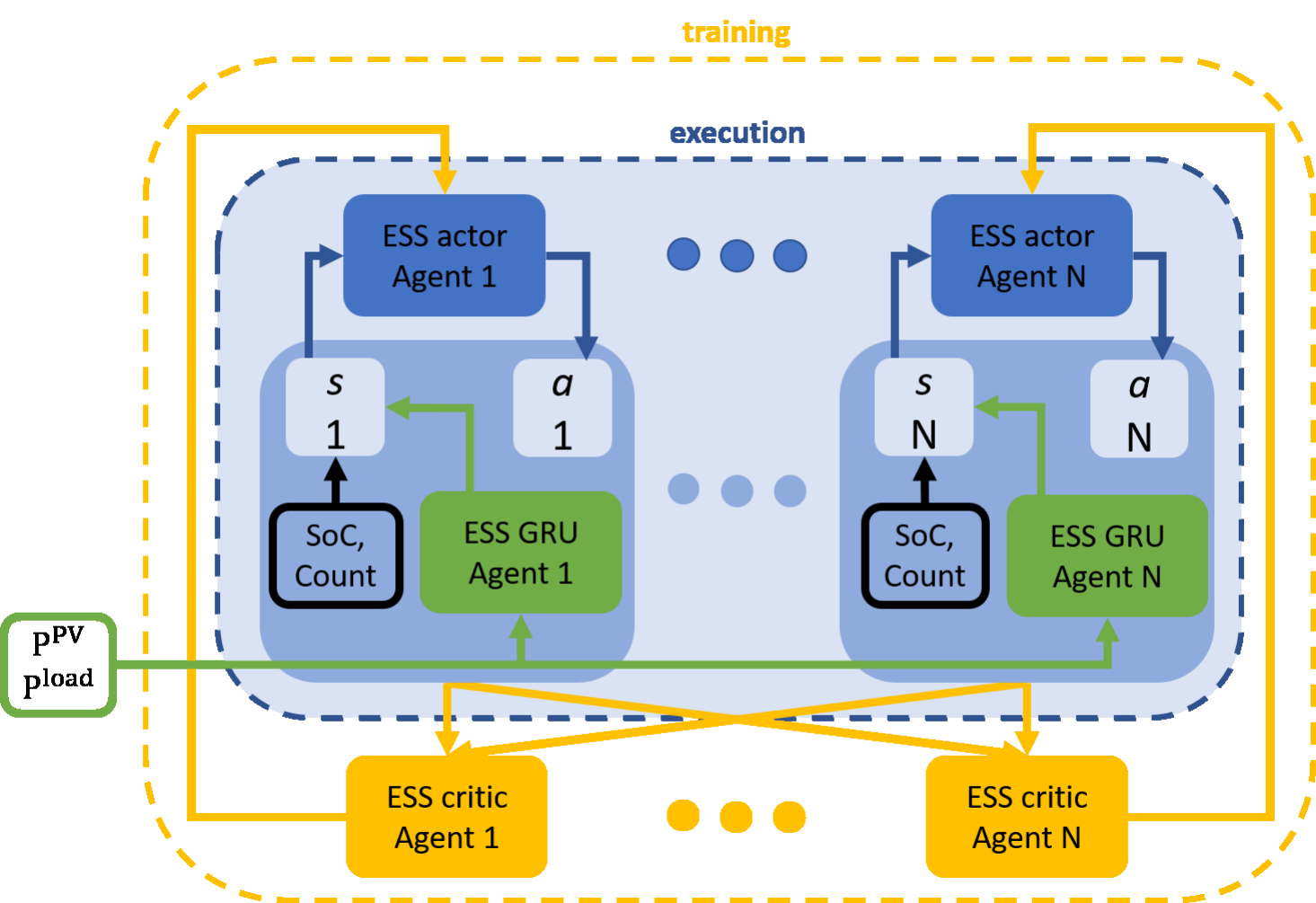}
  \caption{Proposed MADRL structure. The structure consists of multiple ESS  actor agents and corresponding ESS critic agents. Each ESS actor agent contains an ESS GRU and a SoC  Count component, processing state (s) and action (a) information. }\label{fig_framework}
\end{figure}

Fig.~\ref{fig_framework} shows the proposed structure of MADRL. 
Each ESS agent consists of an actor, a GRU, and a critic represented by navy, green, and orange rectangles, respectively, where
$s$ and  $a$ represent the states and actions of the agent, respectively. 
 $N$ agents exist, obtain states, and perform actions in the environment.
During the training process, the ESS actor takes an action based on its state, and then the ESS critic evaluates the action by using all states and actions of the other agents.
This MADDPG-based structure enhances overall performance by encouraging each critic to evaluate all agents’ actions, not just its own.

\subsection{Multiagent Deep  Reinforcement Learning}

Multiagent RL has multiple agents interacting with the environment at the same time.
Each agent observes its own state $s$ and performs action $a$, and the environment sends back reward $r$ to every agent and then proceeds to the next state.
Agents learn to optimize their own policy $\pi_n(s)$ so as to maximize a cumulative reward.

In our scenario, we refer to the $n$th ESS as agent $n$, and its state includes the current state of charge $SoC_{n,t}$, 
a counter $c_{n,t}$ indicating how many time slots are left to meet the highest probability of grid disconnection, and the 
characteristic vector of temporal information $\mathbf{v}_t$. In practice, the counter $c_{n,t}$ may be obtained using a physical prediction model. For instance, a wind speed-based failure curve can be generated based on wind speed prediction \cite{17Panteli}. The failure curve can be further applied to estimate the probability of power disconnection over time \cite{21zhou}.
The state $s_{n,t}$ can thus be expressed as
\begin{equation}\label{eq states}
s_{n,t} = [SoC_{n,t}, c_{n,t}, \mathbf{v}_t].
\end{equation}
A compact state vector can then be constructed by
\begin{equation}\label{eq_all_states}
\mathbf{s}_{t} = [s_{1,t}, s_{2,t}, \cdots, s_{n,t}].
\end{equation}

Let $a_{n,t}$ denote the action of agent $n$, which is designed as the charging/discharging power of ESS $P_{n,t}^{\text{ESS}}$:
\begin{equation}\label{eq actions}
a_{n,t} = [P_{n,t}^{\text{ESS}}].
\end{equation}

The charging/discharging power of ESS depends on the PV generations and power load, suitable for a learning method that can address system uncertainties.
To reduce the action space for faster learning, the power output of generators and load shedding proportion are determined directly. While this simplification may slightly reduce solution optimality, it enables quicker model convergence and easier implementation, potentially offering a favourable trade-off between accuracy and efficiency in many practical applications \cite{le2022hybrid}. The power output of generators $P_{k,t}^{\text{gen}}$ is determined based on the current load demands $P_{l,t}^{\text{load}}$, which can be expressed as
\begin{equation}\label{eq_cal_gen_power}
\begin{split}
P_{k,t}^{\text{gen}} = 
\left\{
  \begin{array}{ll}
P_{k,\text{max}}^{\text{gen}},  \hbox{ if } \displaystyle \sum_{k\in \mathcal{K}}P_{k,\text{max}}^{\text{gen}} \leq \sum_{l\in \mathcal{L}}P_{l,t}^{\text{load}};    \\
\displaystyle \sum_{l\in \mathcal{L}}P_{l,t}^{\text{load}}\frac{P_{k,\text{max}}^{\text{gen}}}{\displaystyle\sum_{k\in \mathcal{K}}P_{k,\text{max}}^{\text{gen}}}, \quad \hbox{ otherwise. }
  \end{array}
\right.
\end{split}
\end{equation}

Load shedding is inevitable to maintain a power balance when the total power generation and storage discharge are less than the total power demand during power outages. The proportion of shedding load is determined by calculating the difference between the total power generation and total power demand. The total power generation during power outages consists of the generation from the PV, generators and ESS discharging ($P_{n,t}^{\text{ESS}} < 0$), and the total power demand consists of load and ESS charging ($P_{n,t}^{\text{ESS}} > 0$). The load shedding proportion $\alpha_{l,t}$ can thus be expressed as
\begin{equation}\label{eq_cal_load_shed}
\alpha_{l,t} = \frac{\displaystyle \sum_{l\in \mathcal{L}}P_{l,t}^{\text{load}} + \sum_{j\in \mathcal{J}}P_{j,t}^{\text{ESS}} - \sum_{i\in \mathcal{I}}P_{i,t}^{\text{PV}} - \sum_{k\in \mathcal{K}}P_{k,t}^{\text{gen}}}{\displaystyle\sum_{l\in \mathcal{L}}P_{l,t}^{\text{load}}}.
\end{equation} 

The fixed generator output strategy adopted may be unable to leverage the dynamic regulation potential of generator units. In scenarios with complex startup and shutdown constraints, this fixed allocation strategy may not achieve an optimal energy distribution, resulting in increased operating costs or reduced system reliability. Additionally, the simplified load shedding model improves computational efficiency but disregards the different priorities of loads, which limits the approach's ability to provide differentiated protection strategies for critical loads, probably resulting in negative effects like economic losses in practical applications.

The main goal of an agent is to maximize the cumulative reward, i.e., minimizing the operational cost in our case.
Based on the cost function in~(\ref{eq cost cal}), 
the reward (or negative cost) can be expressed as
\begin{equation}\label{eq rewards}
\begin{split}
&r_{n,t}=-[\lambda^{\text{ESS}}\left\vert\mathop{\min}\{P_{n,t}^{\text{ESS}},0\}\right\vert+
\sum_{k\in \mathcal{K}}\lambda^{\text{gen}}P_{k,t}^{\text{gen}}\\
&\qquad +\lambda^{\text{grid}} \left\vert P_{t}^{\text{grid}}\right\vert+
\sum_{l\in \mathcal{L}}\alpha_{l,t}\lambda^{\text{load}}P_{l,t}^{\text{load}}] \Delta t.
\end{split}
\end{equation}

\begin{figure}
\centering
  \includegraphics[width=8cm]{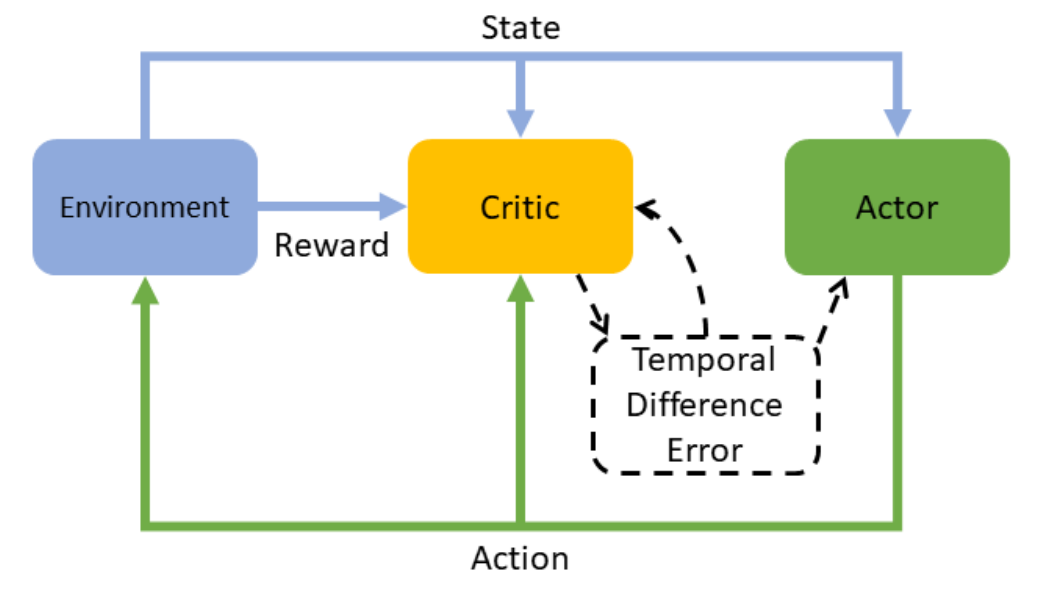}
  \caption{Actor-Critic reinforcement learning structure. It uses an Actor to choose actions based on states, evaluated by a Critic using rewards from the Environment. The Critic's feedback guides updates to both the Actor's policy and the Critic's value function for continual performance improvement.}\label{fig_AC_framework}
\end{figure}

MADDPG is one of the most popular multiagent RL algorithms. The algorithm adopts an actor-critic framework. The actor selects an action based on the observed states from the environment, and the critic evaluates the action by computing a temporal difference error. Fig.~\ref{fig_AC_framework} shows the actor-critic structure. In multiagent configuration, each agent contains two networks: actor and critic networks.
The actor network is a policy network that uses the states of the agent as input and produces an action as the output.
The critic network exploits all states and actions from all the agents to generate state-action values, which can be used to evaluate the current action.
To make the learning process stable, target actor and target critic networks having the same structure as the original networks are used; the only difference is that the parameters of target networks are updated with the parameters of the original networks multiplied by a small value. In this manner, the target networks slowly track the changes in original networks over time \cite{22du}.

Unlike the conventional actor-critic structure, the critic network uses information from other agents.
The critic takes states and actions from all the agents as the input to evaluate the action from agent $n$.
The target value $y_{n,t}$ is the sum of current reward $r_{n,t}$ and target critic network output at the next state.
The loss function $L_{n}(\theta_{n}^{Q})$ is the mean square error between the target value and output from the behaviour critic network, which measures the difference between the behavior critic network and the target critic network.
The target value and mean square error are calculated as
\begin{equation}\label{eq target_val}
\begin{aligned}
y_{n,t}^{(b)} = r_{n,t}^{(b)} + &\gamma Q_{n}^{\prime}(s_{1,t+1}^{(b)},\cdots,s_{N,t+1}^{(b)}, \\
&\pi_{1}(s_{1,t+1}^{(b)};\theta_{1}^{\pi^{\prime}}),\cdots,\pi_{N}(s_{N,t+1}^{(b)};\theta_{N}^{\pi^{\prime}}); \theta_{n}^{Q^{\prime}})
\end{aligned}
\end{equation}

\begin{equation}\label{eq critic_loss}
\begin{aligned}
L_{n}(\theta_{n}^{Q}) = \frac{1}{B} \displaystyle \sum_{b=1}^{B} (y_{n,t}^{(b)} - Q_{n}(&s_{1,t}^{(b)},\cdots,s_{N,t}^{(b)},\\
&a_{1,t}^{(b)},\cdots,a_{N,t}^{(b)}; \theta_{n}^{Q}))^2
\end{aligned}
\end{equation}
where $\gamma$ is the discounting factor,
$\theta_{n}^{Q^{\prime}}$ and $\theta_{n}^{Q}$ represent the parameters of the target critic network and behaviour critic network, respectively,
and $B$ is the sample size, and $b$ is the sample index.

The parameters of the behaviour critic network are updated with learning rate $\eta_{n}^{Q}$ to minimize the mean square error. Next, we update the parameters of the target critic network with a small value $\tau$, which can stabilize the learning process. The update rule can be expressed as
\begin{equation}\label{eq critic_para}
\theta_{n}^{Q} \leftarrow \theta_{n}^{Q} + \eta_{n}^{Q}\nabla L_{n}(\theta_{n}^{Q})
\end{equation}

\begin{equation}\label{eq tar_critic_para}
\theta_{n}^{Q^{\prime}} \leftarrow \tau\theta_{n}^{Q} + (1-\tau)\theta_{n}^{Q^{\prime}}
\end{equation}

The objective of an actor network is to learn an optimal policy $\pi_{n}^{*}(s_{n,t})$ that maximizes the overall reward during the learning process.
First, action is generated by policy $\pi_{n}(s_{n,t}^b;\theta_{n}^{\pi})$, and  the critic network generates the corresponding state-action value.
The gradient of the state-action value is then calculated concerning the parameters of the behaviour actor-network:
\begin{equation}\label{eq actor_updat}
\begin{aligned}
J_{n}(\theta_{n}^{\pi}) = \frac{1}{B} \displaystyle \sum_{b=1}^{B}&Q_{n}(s_{1,t}^{(b)},\cdots,s_{N,t}^{(b)},\\
&\pi_{1}(s_{1,t}^{(b)};\theta_{1}^{\pi}),\cdots,\pi_{N}(s_{N,t}^{(b)};\theta_{N}^{\pi}); \theta_{n}^Q)
\end{aligned}
\end{equation}

\begin{equation}\label{eq actor_grad}
\begin{aligned}
\nabla J_{n}(\theta_{n}^{\pi}) =
\frac{1}{B} \displaystyle \sum_{b=1}^{B} [\nabla &Q_{n}(s_{1,t}^{(b)},\cdots,s_{N,t}^{(b)},\pi_{1}(s_{1,t}^{(b)};\theta_{1}^{\pi}),\cdots,\\
&\pi_{N}(s_{N,t}^{(b)};\theta_{N}^{\pi}); \theta_{n}^Q)\nabla_{\theta_{n}^{\pi}}\pi_{n}(s_{n,t}^{(b)};\theta_{n}^{\pi})]
\end{aligned}
\end{equation}
where $\theta_{n}^{\pi^{\prime}}$ and $\theta_{n}^{\pi}$ are the parameters of the target actor-network and behaviour actor-network, respectively.

Finally, the parameters of the behaviour actor-network and target actor-network 
are updated
in the same manner as critic networks.
The update rule of the actor network can be expressed as
\begin{equation}\label{eq actor_para}
\theta_{n}^{\pi} \leftarrow \theta_{n}^{\pi} + \eta_{n}^{\pi}\nabla J_{n}(\theta_{n}^{\pi})
\end{equation}

\begin{equation}\label{eq tar_actor_para}
\theta_{n}^{\pi^{\prime}} \leftarrow \tau\theta_{n}^{\pi} + (1-\tau)\theta_{n}^{\pi^{\prime}}
\end{equation}
where $\eta_{n}^{\pi}$ represents the learning rate.

\begin{figure}[H]
\centering
  \includegraphics[width=8cm]{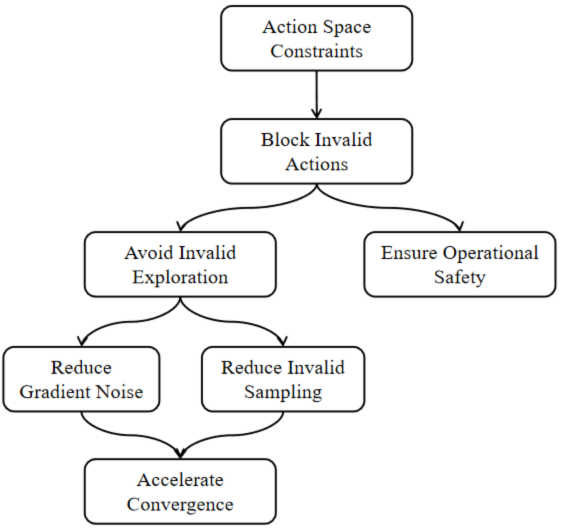}
  \caption{Action masking workflow for accelerating reinforcement learning convergence.}\label{fig_Chain_mask}
\end{figure}

\subsection{Action Masking}
Action masking is a safety mechanism that dynamically limits the action space based on the current SoC of the energy storage system (ESS). When the SoC reaches the upper limit ($SoC \geq SoC_{\max}$), charging commands are masked. Discharging commands are masked when SoC drops to the lower limit ($SoC \leq SoC_{\min}$). The system continuously monitors whether the output of the action by the Actor network would cause the SoC to exceed the safe range. It automatically maps all invalid actions to the nearest feasible boundary. As shown in  Fig.~\ref{fig_Chain_mask}, action masking enhances algorithm performance through a series of mutually reinforcing steps. First, by blocking invalid actions, it avoids invalid exploration while ensuring operational safety and reducing gradient noise. Subsequently, avoiding invalid exploration reduces invalid sampling and shrinks the actual action space. Finally, the combined reduction in gradient noise and invalid sampling accelerates convergence.

In conventional MADDPG, the actions directly come from the output of the actor-network. Although action clipping effectively restricts action values within reasonable bounds to prevent learning instability caused by excessively large or small actions, it can negatively affect learning due to saturation. Therefore, we propose an action masking approach that dynamically adjusts the range of ESS charging/discharging over time based on the current SoC.

\begin{figure}[H]
\centering
  \includegraphics[width=8cm]{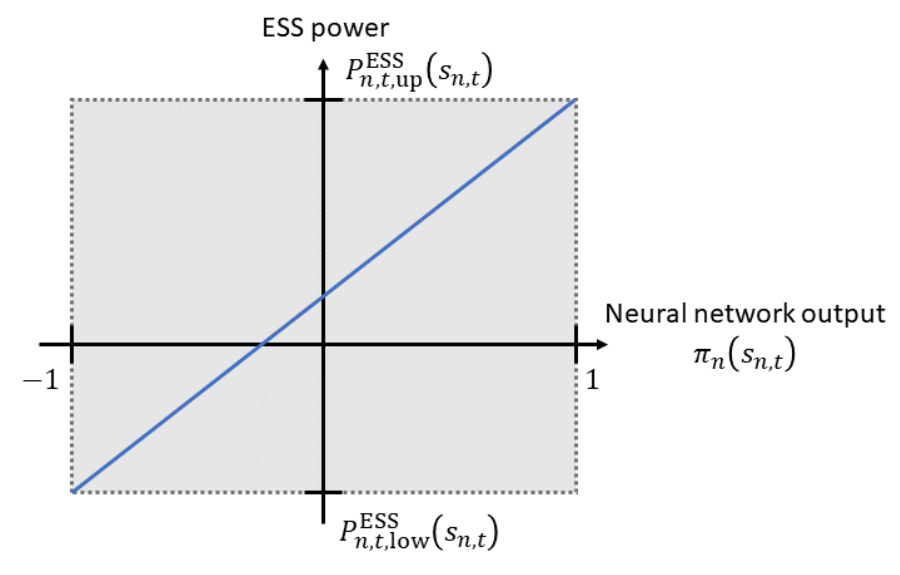}
  \caption{Mapping the neural network output into real ESS power.}\label{fig_action_mask}
\end{figure}

Since the tanh function has an output range of (-1, 1), suppose that the hyperbolic tangent is used as the activation function at the actor output layer; at time $t$, we have $\pi_{n}(s_{n,t})\in(-1,1)$. The tangent function ensures that the computed action variable $a_{n,t}$ is constrained within the feasible range of charging and discharging power capacities. Consequently, this constraint enforcement mechanism guarantees both the operational validity and safety of the energy storage system. The maximum charging power based on the current SoC for the $n\text{th}$ ESS is $P_{n,t,\text{up}}^{\text{ESS}}(s_{n,t})$ and the maximum discharging power is $P_{n,t,\text{low}}^{\text{ESS}}(s_{n,t})$. Fig.~\ref{fig_action_mask} shows the transformation from $\pi_{n}(s_{n,t})$ to ESS power, ensuring that the charging/discharging power control is within a valid range. The calculations for $P_{n,t,\text{up}}^{\text{ESS}}(s_{n,t}), P_{n,t,\text{low}}^{\text{ESS}}(s_{n,t})$ and valid action $a_{n,t}$ can then be expressed as
\begin{equation}\label{fc action_map}
\begin{aligned}
a_{n,t}&= \\
&\frac{\left( P_{n,t,\text{up}}^{\text{ESS}}(s_{n,t})-P_{n,t,\text{low}}^{\text{ESS}}(s_{n,t}) \right)(\pi_{n}(s_{n,t})+1)}{2} \\
&+ P_{n,t,\text{low}}^{\text{ESS}}(s_{n,t})
\end{aligned}
\end{equation}

\begin{equation}\label{eq act_upper}
\begin{aligned}
P_{n,t,\text{up}}^{\text{ESS}}&(s_{n,t})=\\
&\mathop{\min}\left( \frac{(SoC_{n,\text{max}}-SoC_{n,t})E_{n}^{\text{ESS}}}{\Delta t}, P_{n,\text{max}}^{\text{ESS}} \right)
\end{aligned}
\end{equation}

\begin{equation}\label{eq act_lower}
\begin{aligned}
P_{n,t,\text{low}}^{\text{ESS}}&(s_{n,t})= \\
&\mathop{\max}\left( \frac{(SoC_{n,\text{min}}-SoC_{n,t})E_{n}^{\text{ESS}}}{\Delta t}, P_{n,\text{min}}^{\text{ESS}} \right).
\end{aligned}
\end{equation}

Algorithm 1 presents the pseudocode of the proposed MADRL framework for improving the resilience of a microgrid. The input contains the PV generations, power loads, and parameters of ESS, generators.
With the knowledge of which time slot has the highest risk of grid disconnection, each agent schedules its own ESS to keep the operational cost at a minimum during the possible outage.

\begin{algorithm}[H]
	\caption{Proposed MADRL Algorithm for Improving the Resilience of Microgrid}
	\label{alg:energy_management}
	\begin{algorithmic}[1]
		\REQUIRE Initialization of parameters: \\$P_{n,\text{min}}^{\text{ESS}}, P_{n,\text{max}}^{\text{ESS}}, SoC_{n}^{\text{min}}, SoC_{n}^{\text{max}}$, $P_{i}^{\text{PV}}$, $P_{l}^{\text{load}}$, $\hat{P}_{i}^{\text{PV}}$, $\hat{P}_{l}^{\text{load}}$, $P_{k,\text{min}}^{\text{gen}}$, $P_{k,\text{max}}^{\text{gen}}$, $\hat{F}$, $\theta_{n}^{\pi},\theta_{n}^{\pi^{\prime}},\theta_{n}^{Q},\theta_{n}^{Q^{\prime}}$, $\mathcal{D}$
		\ENSURE Energy management policy for improving the resilience of microgrid:
		\FOR{$episode = 0, 1, 2, ..., M$}
		\STATE Initialize state $\mathbf{s}_{0}$
		\FOR{$t = 0, 1, 2, ..., T$}
		\STATE Choose action $a_{n,t}$ according to $s_{n,t}$ and add noise for action exploration for all agents.
		\STATE Perform all actions $(a_{1,t},...,a_{N,t})$, observe rewards $(r_{1,t},...,r_{N,t})$ and the next state $\mathbf{s}^{\prime}_{t}$.
		\STATE Store $(\mathbf{s}_{t}, a_{1,t},...,a_{N,t}, r_{1,t},...,r_{N,t}, \mathbf{s}^{\prime}_{t})$ in experience replay $\mathcal{D}$
		\FOR{each agent}
		\STATE From experience replay, random sample a batch of $(\mathbf{s}_{t}, a_{1,t},...,a_{N,t}, r_{1,t},...,r_{N,t}, \mathbf{s}^{\prime}_{t})$.
		\STATE Update critic according to ~(\ref{eq target_val})--(\ref{eq critic_para}).
		\STATE Update actor according to ~(\ref{eq actor_updat})--(\ref{eq actor_para}).
		\STATE Update target according to ~(\ref{eq tar_critic_para}) and (\ref{eq tar_actor_para}).
		\ENDFOR
		\STATE $\mathbf{s}_{t} \leftarrow \mathbf{s}^{\prime}_{t}$
		\ENDFOR
		\ENDFOR
	\end{algorithmic}
\end{algorithm}

\subsection{Hyperparameter Settings and Network Architecture}
The proposed MADDPG framework included five agents, each controlling an ESS. To ensure stable and efficient training of these agents, a set of experimentally tuned hyperparameters and network structures was adopted, as shown in Tables \ref{tab:train_hyperparams} and \ref{tab:net_arch}. The learning rates of the Actor, Critic, and GRU modules were uniformly set to 0.00025 to ensure consistency in the convergence process. The discount factor was set to 0.99 to emphasize long-term rewards, while the soft update coefficient of 0.001 was used to stabilize the update of the target network. A batch size of 128 was used, and model parameters were updated once every 24 timesteps through a single gradient descent step. An initial exploration phase of 8,000 steps was introduced to encourage exploration before policy learning. The entire training process consisted of 400 episodes.

As shown in Table \ref{tab:net_arch}, the Actor and Critic networks had input dimensions of 18 and 26, respectively. Both networks employed two hidden layers with 64 units each, combined with the LayerNorm normalization technique and ReLU activation function. The GRU module structure comprised a 32-unit input embedding layer, two GRU layers with 32 hidden units each, and a 16-unit output layer activated by ReLU.

\begin{table}[ht]
\caption{Key hyperparameter configurations for the MADDPG framework during training}
\centering
\begin{tabular}{l l}
\hline
\textbf{Parameter} & \textbf{Value} \\
\hline
Actor learning rate & 0.00025 \\
Critic learning rate & 0.00025 \\
GRU learning rate & 0.00025 \\
Discount factor & 0.99 \\
Soft-update coefficient & 0.001 \\
Batch size & 128 \\
Update frequency (timesteps) & 24 \\
Updates per frequency & 1 \\
Warm-up (exploration) steps & 8{,}000 \\
Total training episodes & 400 \\
Number of agents & 5 \\
Action dimensionality (per agent) & 1 \\
\hline
\end{tabular}
\label{tab:train_hyperparams}
\end{table}

\setlength{\intextsep}{10pt}
\begin{figure}[H]
\centering
  \includegraphics[width=8.5cm]{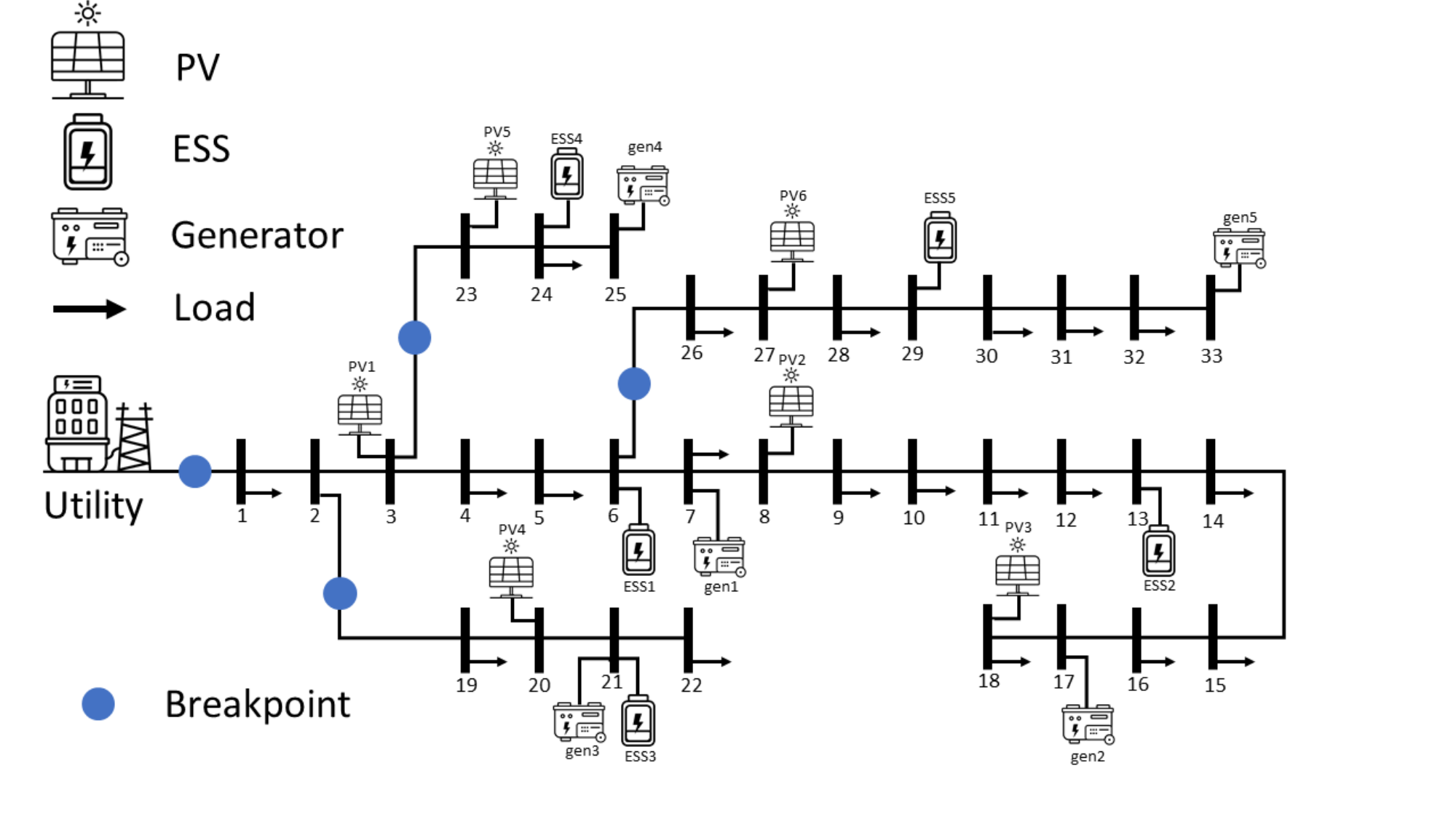}\\
  \caption{IEEE 33-bus Distribution System with Integrated Distributed Energy Resources (DERs): A Schematic Representation of PV, ESS, and Generator Integration}\label{fig_bus_sys}
\end{figure}

\setlength{\intextsep}{6pt}

\section{Numerical Results}\label{sec_result}
This section examined energy management in a microgrid by using the proposed method. One time slot length was set to 15 minutes.
The IEEE 33-bus system \cite{89baran}  in Fig.~\ref{fig_bus_sys}
was investigated, with five ESS, six PV systems, five generators, and twenty loads specified in
Tables~\ref{ESS param},~\ref{DER param} and~\ref{load param}.
Pandapower \cite{18thurner} was used to perform alternating current flow analysis. 
The following parameters were set: ESS charging and discharging efficiency $\delta^{\text{ch}}=0.999$ and $\delta^{\text{dis}}=1.001$ in~(\ref{eq ESS SoC dynamic}) were set, degradation cost $\lambda^{\text{ESS}}=$ \$0.2/MWh, generators cost $\lambda^{\text{gen}}=$ \$0.5/MWh, electric price $\lambda^{\text{grid}}=$ \$0.3/MWh, and load shedding cost $\lambda^{\text{load}}=$ \$1.5/MWh.
The observation window ($T$) for GRU was 8-time slots (120 minutes).
Our algorithm parameters were manually tuned for best performance. The discount rate $\gamma = 0.99$ was set to reflect the importance of the long-term return. To avoid unstable learning or slow convergence, we set actor and critic learning rates $\eta^{\pi} = \eta^{Q} = 2.5 \times 10^{-4}$. The target network soft update rate $\tau = 0.001$ was set to improve stability.

\begin{figure} [H]
\begin{equation*}
\begin{array}{c}
  \includegraphics[width=8cm]{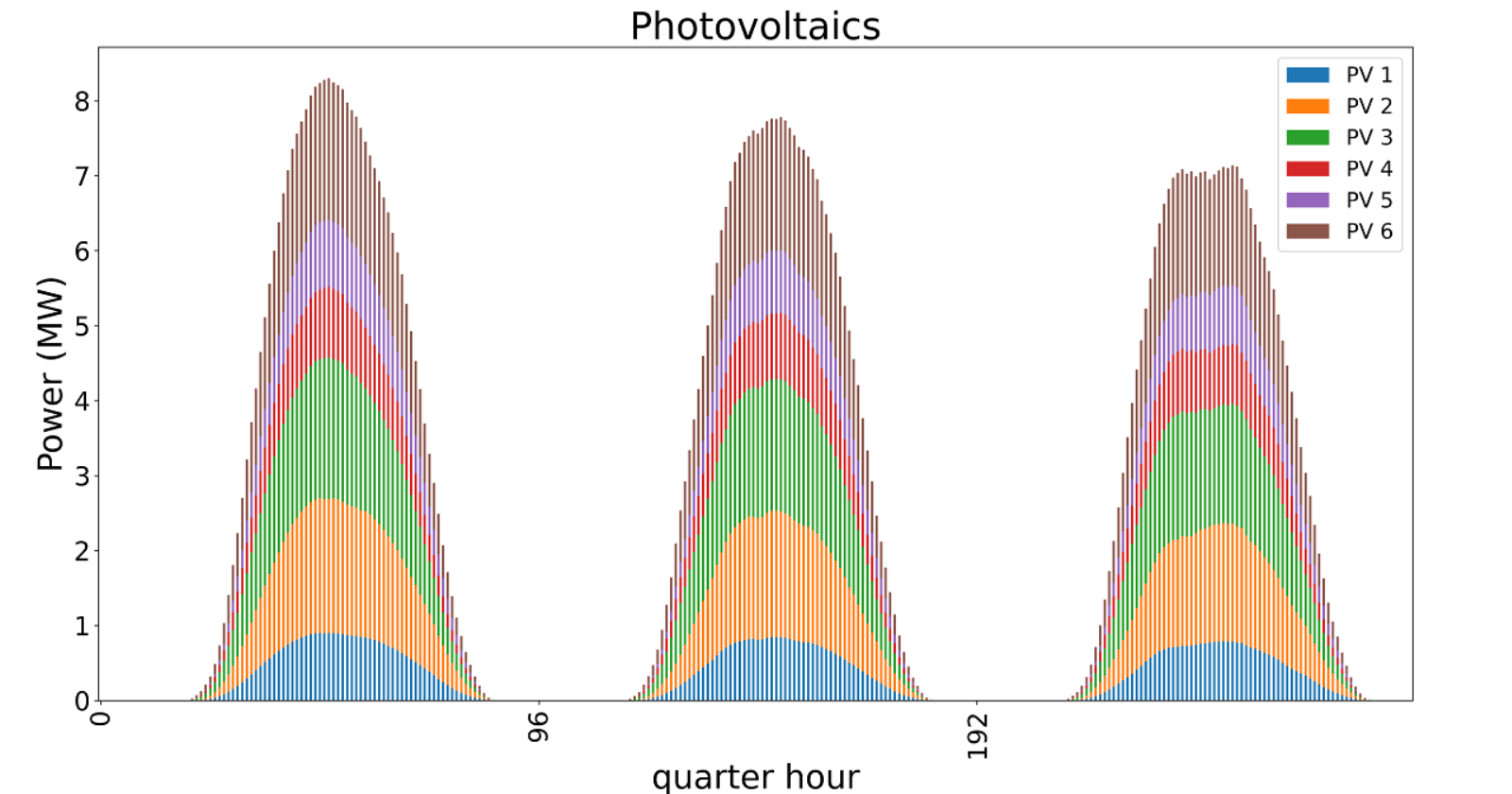}\\
  \mbox{(a)}\\
  \includegraphics[width=8cm]{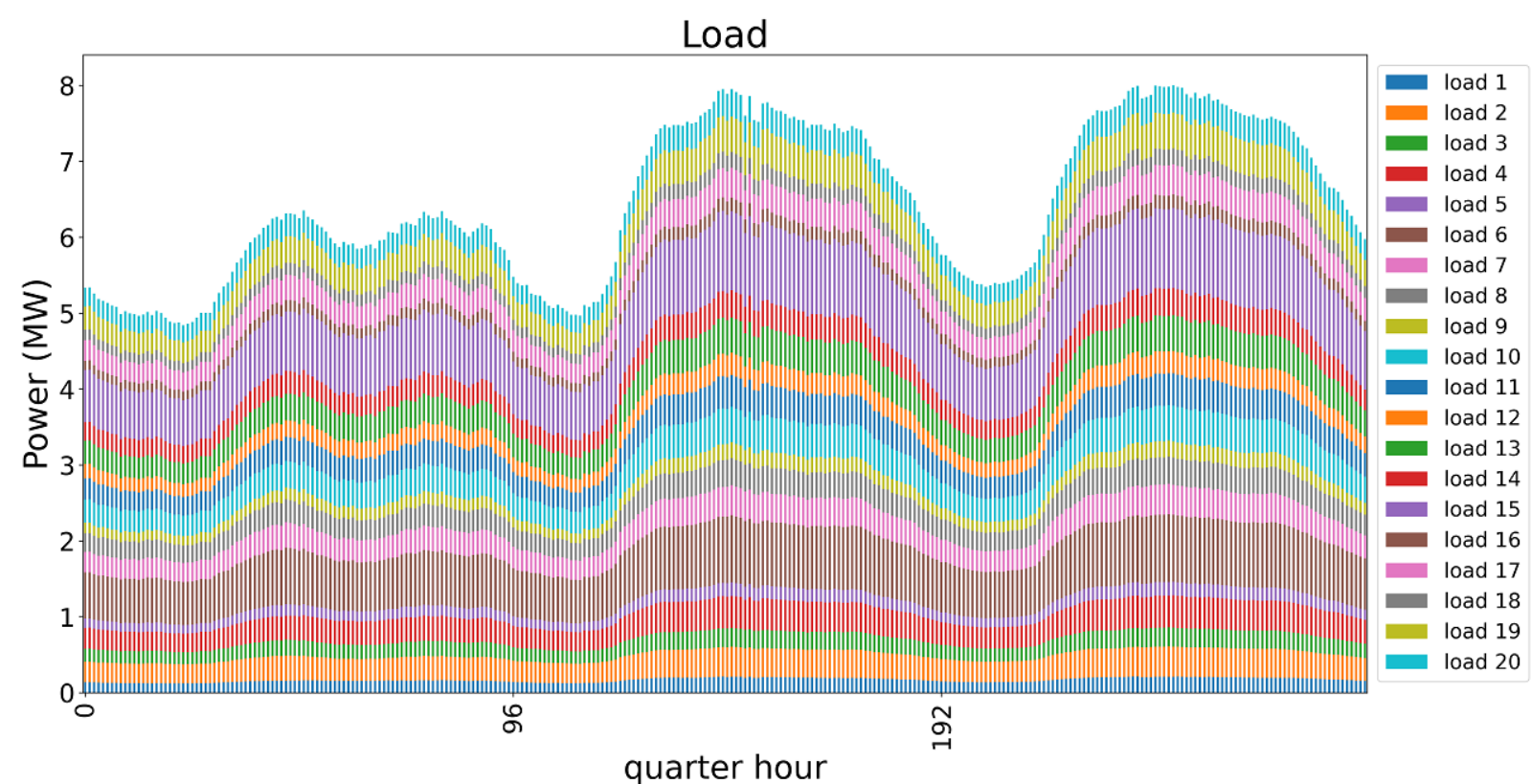}\\
  \mbox{(b)}\\
  \includegraphics[width=8cm]{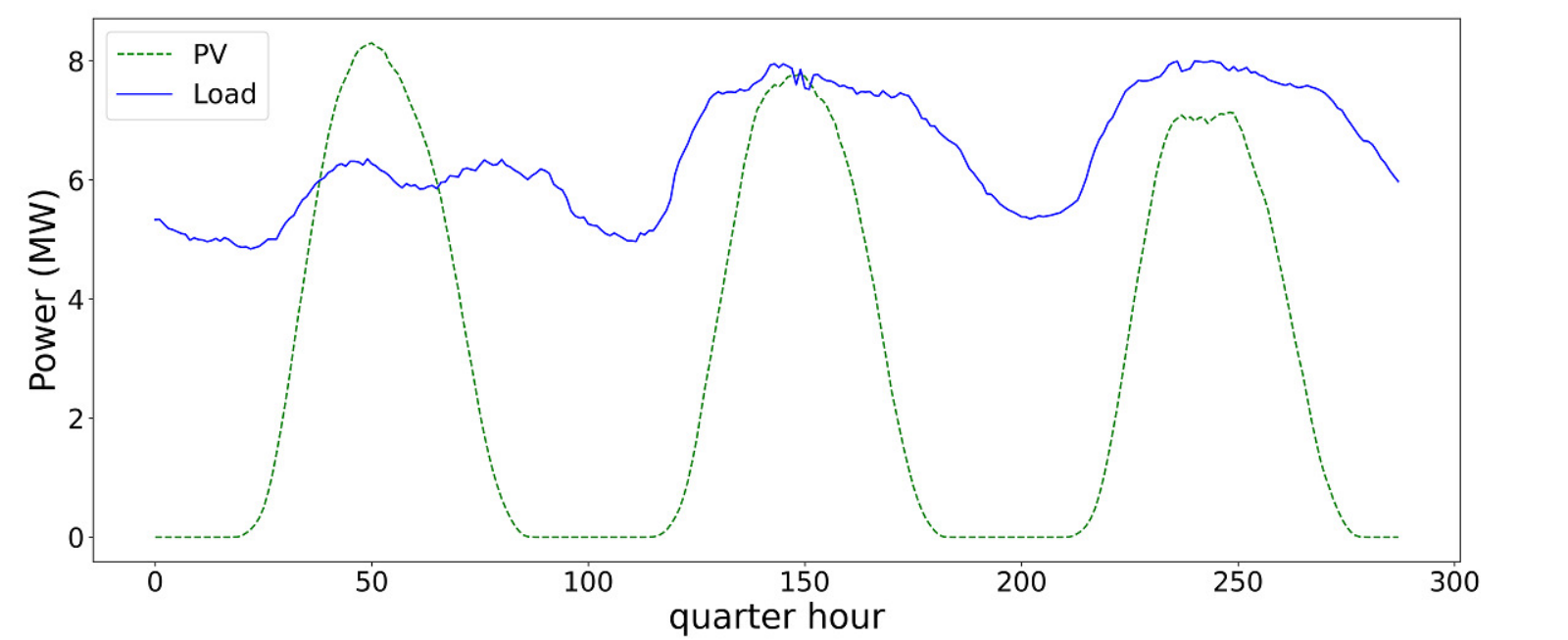}\\
  \mbox{(c)}  
\end{array}
\end{equation*}    
\caption{Samples of real-world data for (a) PV generation (b) power load, and (c) aggregate PV generation and power load.}\label{fig_PV_load_data}
\end{figure}

Real-world data about PV generation and power load in July and August 2022 were collected from the European Network of Transmission System Operators for Electricity (ENTSO-E) \cite{Web1}.
A quarter-hourly resolution was considered.
Data was scaled to fit the value ranges defined in Tables~\ref{DER param} and~\ref{load param}
and fed into the IEEE 33-bus system;
Fig.~\ref{fig_PV_load_data} shows some samples from the database.
We modeled prediction errors for load and PV generation using Gaussian distributions, incorporating uncertainty in the simulation.

\begin{table}[ht]
\caption{Network architecture for MADDPG component}
\centering
\begin{tabular}{l l}
\hline
\textbf{Component} & \textbf{Architecture} \\
\hline
\multicolumn{2}{l}{\textbf{Actor Network}} \\
Input layer & State dimension (18)  \\
Dense (fc1) & 64 units, LayerNorm + ReLU \\
Dense (fc2) & 64 units, LayerNorm + ReLU \\
Output layer & 1 unit (per agent), tanh \\
\hline
\multicolumn{2}{l}{\textbf{Critic Network}} \\
Input layer & Global state dimension (26) \\
Dense (fc1) & 64 units, LayerNorm + ReLU \\
Dense (fc2) & 64 units, LayerNorm + ReLU \\
Action embedding & 64 units, ReLU \\
Output layer & 1 unit, Linear \\
\hline
\multicolumn{2}{l}{\textbf{GRU Module}} \\
Input embedding & 32 units, ReLU \\
GRU layers & 2 layers, hidden size 32 \\
Output layer & 16 units, Linear + ReLU \\
\hline
\end{tabular}
\label{tab:net_arch}
\end{table}

\begin{table}
\centering
\caption{Parameters of Energy Storage System}
\label{ESS param}
\resizebox{\linewidth}{!}{%
\begin{tabular}{cccccc} 
\hline
Device ID & \begin{tabular}[c]{@{}c@{}}Minimum Power\\MW\end{tabular} & \begin{tabular}[c]{@{}c@{}}Maximum Power\\MW\end{tabular} & \begin{tabular}[c]{@{}c@{}}Maximum Energy\\MWh\end{tabular} & Minimum SoC & Maximum SoC \\ 
\hline
$\text{ESS}_1$ & -2 & 2 & 6 & 0.1 & 0.9 \\
$\text{ESS}_2$ & -1.5 & 1.5 & 4 & 0.1 & 0.9 \\
$\text{ESS}_3$ & -2 & 2 & 6 & 0.1 & 0.9 \\
$\text{ESS}_4$ & -1 & 1 & 3 & 0.1 & 0.9 \\
$\text{ESS}_5$ & -1 & 1 & 3 & 0.1 & 0.9 \\
\hline
\end{tabular}
}
\end{table}

\begin{table}[H]
\centering
\caption{Parameters of Generators and Photovoltaic Modules}
\label{DER param}
\resizebox{\linewidth}{!}{%
\begin{tabular}{ccc|ccc} 
\hline
Gen ID & \begin{tabular}[c]{@{}c@{}}Minimum Power\\MW\end{tabular} & \begin{tabular}[c]{@{}c@{}}Maximum Power\\MW\end{tabular} & PV ID & \begin{tabular}[c]{@{}c@{}}Minimum Power\\MW\end{tabular} & \begin{tabular}[c]{@{}c@{}}Maximum Power\\MW\end{tabular} \\ 
\hline
$\text{Gen}_1$ & 0 & 2 & $\text{PV}_1$ & 0 & 1 \\
$\text{Gen}_2$ & 0 & 1 & $\text{PV}_2$ & 0 & 2 \\
$\text{Gen}_3$ & 0 & 1 & $\text{PV}_3$ & 0 & 2 \\
$\text{Gen}_4$ & 0 & 1 & $\text{PV}_4$ & 0 & 1 \\
$\text{Gen}_5$ & 0 & 1 & $\text{PV}_5$ & 0 & 1 \\
- & - & - & $\text{PV}_6$ & 0 & 2 \\
\hline
\end{tabular}
}
\end{table}

\begin{table}
\centering
\caption{Amount of load shedding for different $\lambda^{\text{load}}$}
\label{diff_load_cost}
\resizebox{\linewidth}{!}{%
\begin{tabular}{ccc} 
\hline
$\lambda^{\text{load}}$ & \begin{tabular}[c]{@{}c@{}}Load Shedding Cost\\MWh~ ~\end{tabular} & \begin{tabular}[c]{@{}c@{}}Avg. Load Shedding / day\\MW\end{tabular} \\ 
\hline
1 & \$ 0.15 & 3.4 \\
2 & \$ 1.5 & 3.15 \\
3 & \$ 30 & 2.84 \\
\hline
\end{tabular}
}
\end{table}

\begin{figure}[H]
\centering
  \includegraphics[width=8cm]{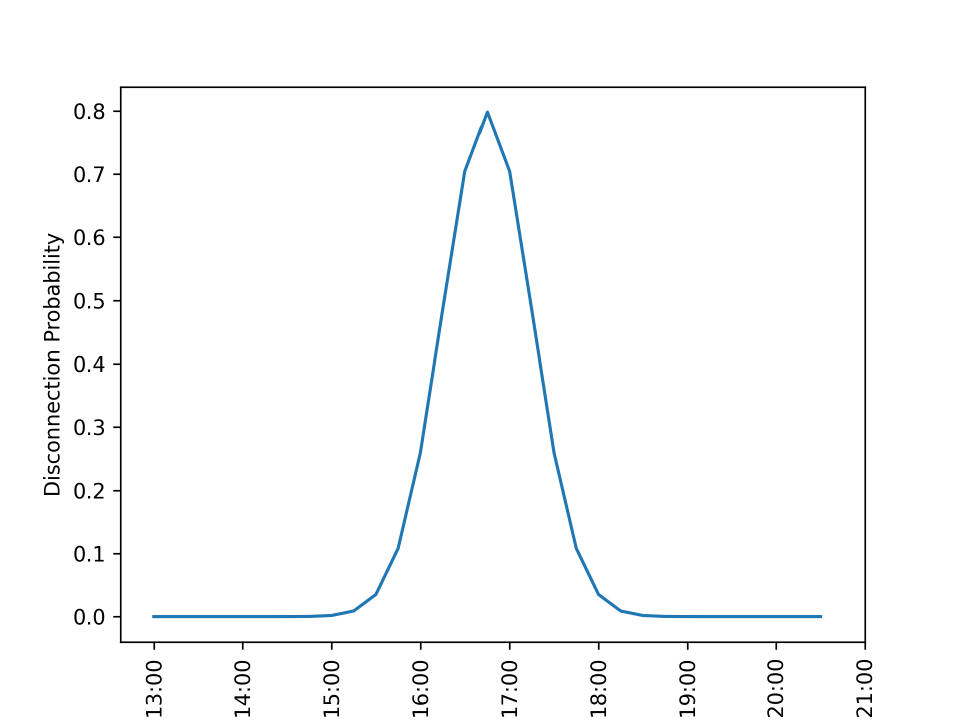}\\
  \caption{Disconnection probabilities in one day.}\label{fig_disc_prob}
\end{figure}

In order to model the effect of a random stormy day, a bell-shaped probability distribution was used.
An example is shown in Fig.~\ref{fig_disc_prob}.
In each episode (day), a time slot was randomly chosen to have the highest probability of 
disconnecting from the main grid, and other disconnection probabilities were symmetric about the chosen time slot. 
A random number between 0 and 1 was generated in each time slot, and the disconnection happened if this number was smaller than the current disconnection probability. 
The duration of the power outage was randomly set between 3 and 4 hours (lasting for 12--15 time slots).
The other 3 breakpoints had similar disconnection probabilities except that the time slots having the highest probability were randomly shifted between -45 and 45 minutes, leading to a difference of 3 time slots.

\begin{table}
\centering
\caption{Parameters of Loads}
\label{load param}
\resizebox{\linewidth}{!}{%
\begin{tabular}{cc|cc|cc} 
\hline
Device ID       & \begin{tabular}[c]{@{}c@{}}Max Power\\MW\end{tabular} & Device ID        & \begin{tabular}[c]{@{}c@{}}Max Power\\MW\end{tabular} & Device ID        & \begin{tabular}[c]{@{}c@{}}Max Power\\MW\end{tabular}  \\ 
\hline
$\text{Load}_1$ & 0.23                                                  & $\text{Load}_8$  & 0.46                                                  & $\text{Load}_{15}$ & 1.14                                                   \\
$\text{Load}_2$ & 0.51                                                  & $\text{Load}_9$  & 0.23                                                  & $\text{Load}_{16}$ & 0.23                                                   \\
$\text{Load}_3$ & 0.32                                                  & $\text{Load}_{10}$ & 0.51                                                  & $\text{Load}_{17}$ & 0.51                                                   \\
$\text{Load}_4$ & 0.46                                                  & $\text{Load}_{11}$ & 0.46                                                  & $\text{Load}_{18}$ & 0.23                                                   \\
$\text{Load}_5$ & 0.23                                                  & $\text{Load}_{12}$ & 0.32                                                  & $\text{Load}_{19}$ & 0.51                                                   \\
$\text{Load}_6$ & 1.14                                                  & $\text{Load}_{13}$ & 0.51                                                  & $\text{Load}_{20}$ & 0.46                                                   \\
$\text{Load}_7$ & 0.51                                                  & $\text{Load}_{14}$ & 0.46                                                  & -                & -                                                      \\
\hline
\end{tabular}
}
\end{table}

\setlength{\intextsep}{1pt}

\subsection{Training Result}
We divided the dataset into training sets (75\%) and testing sets (25\%). In Table~\ref{diff_load_cost}, setting a higher load shedding cost $\lambda^{\text{load}}$ in the proposed method yielded a smaller amount of load shedding, which indicates that adjusting load shedding cost $\lambda^{\text{load}}$ can reduce the amount of load being unserved.

\begin{table*}[t]
  \centering
  \caption{Testing results comparison of different approaches}
  \label{test_res}
  \resizebox{\textwidth}{!}{%
  \begin{tabular}{l c c c c c}
    \hline
    \textbf{Method} & \textbf{Avg. cost (\$)} & \textbf{Highest cost (\$)} & \textbf{Lowest cost (\$)} & \textbf{Avg. load shedding (MW)} & \textbf{Computation time (s)} \\
    \hline
    Rule-based      & 41.51 & 46.42 & 31.69 & \textbf{\textasciitilde{}~0} & \textbf{513.67} \\
    MINLP           & \textbf{35.06} & 45.08 & \textbf{25.15} & 4.12 & 3{,}632.48 \\
    DDPG            & 37.71 & 44.67 & 28.54 & 2.94 & 527.07 \\
    MAPPO           & 35.64 & 51.07 & 27.87 & 0.96 & 966.20 \\
    Proposed method & 35.18 & \textbf{42.17} & 25.83 & 1.72 & 529.63 \\
    \hline
  \end{tabular}
  }
\end{table*}

\begin{table*}[t]
  \centering
  \caption{Testing results under different agents failure scenarios}
  \label{tab:failure_scenarios}
  \resizebox{\textwidth}{!}{%
  \begin{tabular}{l c c c c}
    \hline
    \textbf{Scenarios} & \textbf{Avg. cost (\$)} & \textbf{Highest cost (\$)} & \textbf{Lowest cost (\$)} & \textbf{Avg. load shedding (MW)} \\
    \hline
    Without failure     & 35.18 & 42.17 & 25.83 & 1.72 \\
    1 agent failure     & 37.16 & 42.63 & 26.63 & 1.57 \\
    3 agents failure    & 37.95 & 42.81 & 26.65 & 1.55 \\
    \hline
  \end{tabular}
  }
\end{table*}

The proposed method was compared with DDPG \cite{19yang}, MINLP \cite{18beal}, and a rule-based method modified from \cite{20sun}. Additionally, MAPPO \cite{wang2024scalable}  was also included as a baseline method to evaluate the effectiveness of MARL approaches in handling the cooperative control of multiple ESS. Two hours (8 time slots) of current and prediction data about PV generation and power load were utilized by the proposed method, DDPG, MAPPO, and MINLP.
The modified rule-based method only used information about the current PV generation and power load because its mechanism was based on the SoC of ESS. Maintaining the SoC around 0.5 balances the charging and discharging flexibility of ESS, ensuring sufficient emergency power reserves.

\begin{figure}[H]
\centering  
  \includegraphics[width=8cm]{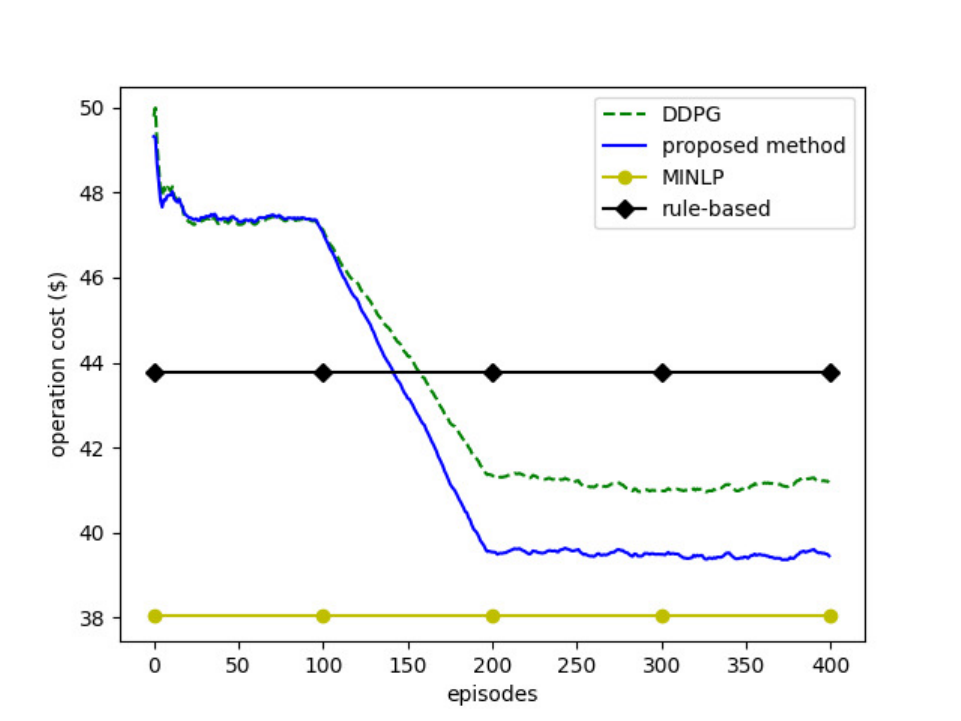}\\
\caption{Learning curves of comparable methods given power outages caused by natural disasters.}\label{fig learning_curve_compa}
\end{figure}

Fig.~\ref{fig learning_curve_compa} shows the associated operational costs.  
We trained the proposed method and DDPG, and both methods converged after 200 episodes.
The proposed method had a lower operational cost than DDPG and allowed for collaboration between multiple agents 
to provide efficient energy management. 
By contrast, conventional DDPG controlled all ESS by one agent, yielding some performance degradation.

The MINLP had a low operational cost because learning-based methods would consider the possibilities of power line disconnection and thus store more energy to prepare for an outage that might not happen. As a result, additional operational costs of learning-based methods were incurred as compared with MINLP.
The rule-based method took the safest way by maintaining the SoC around 0.5 without considering the probabilities of disconnection, yielding the highest operational costs.

\subsection{Testing Result}
Data from 16 days were used to test the proposed method, DDPG, rule-based, MINLP, as well as MAPPO. As shown in Table~\ref{test_res}, the testing results of the proposed method are compared with several mainstream baseline methods in the cost, amount of load shedding and computation time of straining. Although MINLP had the lowest average cost, it did not ensure emergency operations during power outages. As a result, the ESS barely supported load demand during power outages; its average load shedding reached 4.12 MW/day, significantly higher than other methods.

The rule-based method nearly had no load shedding, but the operational costs were the highest, averaging \$41.51 per day. This was because the SoC of the ESS was maintained at 0.5 regardless of power outages. By contrast, the proposed method efficiently coordinated ESS charging/discharging based on the predictions of power load and PV generation. Before the outages happened, the ESS were charged precautionary. Consequently, the highest cost and average load shedding of the proposed method were both better than MINLP, at \$42.17 per day and 1.72 MW per day, respectively.

In addition, the operational cost and load shedding of DDPG were still worse than the proposed method. This is because the proposed method allows multiple agents to control ESS collaboratively, while DDPG controls all ESS by one agent. Notably, MAPPO achieved an average operational cost of \$35.64, which was between the MINLP and our proposed method. MAPPO reduced average load shedding to 0.96 MW per day, which demonstrates its effectiveness in multiple agents under weather uncertainty. However, this performance gain comes with increased computational expense, which is much higher than 529.63 s for our proposed framework. Overall, the proposed method demonstrated a better balance between operational cost, load shedding and computational efficiency.

To further validate the robustness of the proposed approach under inaccurate PV generation and load forecasting conditions, we simulated prediction deviations where actual PV generation decreased by 15\% and load increased by 15\% compared to predicted values. The results demonstrate that operating costs increased as expected, with the most significant change occurring in the minimum cost scenario, which increased by 6.35\% to \$27.47 per day. However, load shedding remained at a relatively low level of 2.77 MW per day.
Moreover, the proposed framework was also evaluated under agent failure scenarios. As shown in Table~\ref{tab:failure_scenarios}, the average cost increased slightly from \$35.18 to \$37.95, while average load shedding decreased to 1.55 MW when 3 agents failed. These results demonstrated that the proposed framework is robust to partial agent failures with only minor performance decline. This minor cost increase is due to the decentralized policy's ability to adapt locally despite partial communication loss, allowing the system to preserve global performance through redundant decision pathways.

\begin{figure}[H]
\centering
  \includegraphics[width=8cm]{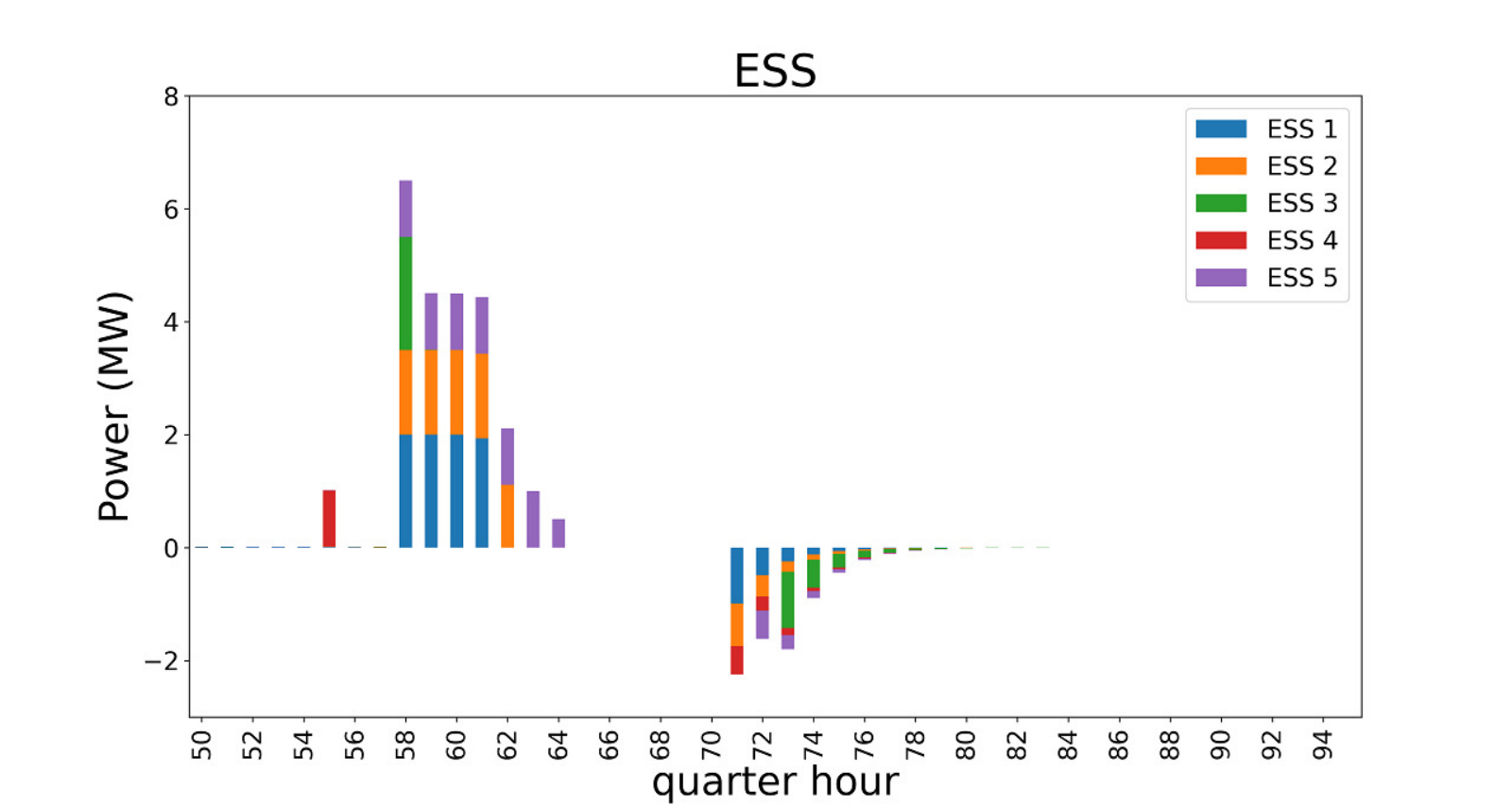}\\
  \caption{ESS charging and discharging of the proposed method to address a power outage.}\label{fig ESS_control}
\end{figure}

Finally, Fig.~\ref{fig ESS_control} presents the ESS charging/discharging profile of the proposed method.
Aware of the potential disconnection from the main grid, the EMS proactively charged the ESS as a precaution.
When a power outage occurred, ESS was discharged to meet the power load as much as possible.
This illustrates the effectiveness of charging/discharging control of our MADRL framework.

This research is primarily based on an IEEE 33-bus radial distribution system, which demonstrates excellent performance with clear learning and convergence trends. This is mainly because radial topologies have relatively simple power flow patterns and single power transmission paths, which result in a lower complexity of the multi-agent state and action spaces, and enabling agents to learn effective coordination strategies more quickly \cite{hu2022multi}.

However, to investigate the performance of our proposed method on different microgrid topologies, we also introduced an IEEE 14-bus meshed system for comparative validation. The results show that the proposed MADRL method's performance is limited in meshed topologies, with the highest operating cost approximately 5.7 times that of the IEEE 33-bus system, reaching \$241.9. Additionally, the computational complexity is also significantly higher than the results in the IEEE 33-bus system. This is because meshed topologies introduce multiple power flow paths and complex interdependencies between nodes that greatly increase the learning complexity \cite{qiu2022coordination}. Multi-agents need to process more state information, making effective coordination exceptionally difficult.

Furthermore, the number and placement of DERs are also critical factors affecting the performance of MADRL-based EMS. Therefore, future work will focus on addressing cross-topology generalization issues by developing more robust algorithm architectures and training strategies to maintain good performance across various microgrid configurations.

\section{Conclusion and Future Work}\label{sec_conclu}
Research on improving the resilience of microgrids has gained much attention because
catastrophic disasters have occurred more frequently in recent years.
A well-designed EMS can manage DERs efficiently and reliably, and further enhance the resilience of a microgrid. To control multiple DERs, we developed a MADRL framework for microgrid energy management. The goal was to improve microgrid resilience by minimizing its operational cost.
We introduced a recursive temporal data extraction model and action masking mechanism to optimize ESS charging/discharging ranges. Numerical analysis using real-world PV generation and power load was conducted, and the uncertainty of power outages was analysed.
We developed mechanisms such as the recursive model for temporal data extraction and action masking for an efficient ESS charging/discharging range. Numerical analysis using real-world PV generation and power load was conducted, and the uncertainty of power outages was analysed.
The proposed MADRL was compared with single-agent DDPG, MINLP, and a modified rule-based method. 
Our simulations show that MINLP was not suitable when power outages occurred opportunistically. 
Although the rule-based method can endure outages, the operational cost was higher because it did not utilize temporal information well.
The proposed MADRL method outperformed comparable methods because it allowed multiple agents
to collaboratively learn their policies for ESS control. 

Although this study has been focused on enhancing the resilience of a single microgrid, it is worth noting that multi-microgrid systems hold significant potential for responding to extreme weather events. Our future work could extend to scenarios involving multiple microgrids, which will present several implementation challenges. For example, 5G wireless systems are seen as promising communication methods for multi-microgrid configurations, but they may raise privacy issues and require the use of energy-intensive antennas. The high energy consumption of these antennas further complicates efficient energy management for a multi-microgrid system. The uncertainty associated with renewable energy sources and weather conditions can lead to significant deviations in PV and load prediction data, thereby increasing the complexity of implementing EMS. Consequently, a robust energy management approach that combines real-time control strategies with forecasting methods is essential to ensure optimal performance even when there are substantial discrepancies between predicted and actual data.
Additionally, certain design assumptions still limit the applicability of the proposed model, future research is supposed to consider joint optimization of generator output and load shedding strategies to achieve more flexible and precise energy management. By introducing a load reduction mechanism that considers economic costs and a dynamic optimization design for generator output, the system's generalization performance will be improved in more complex and realistic scenarios. The reactive components in AC lines reduce power transmission capacity due to thermal losses, affecting the frequency stability between microgrids. While DC lines can mitigate stability issues by eliminating reactive power flows, the lack of implementation experience adds uncertainty to their use in multi-microgrid systems. These challenges warrant further research.

\section*{Declaration of generative AI and AI-assisted technologies in the writing process}

During the preparation of this work the authors used ChatGPT in order to polish our writing. After using this tool/service, the authors reviewed and edited the content as needed and take full responsibility for the content of the publication.

\section*{Acknowledgements}
This work has been supported by the Royal Society through 
the project of "CANDlES: Consumer-Centric Analytics for Net-Zero Digitalised Multi-Energy Systems" (KTP\textbackslash R1\textbackslash 241042). We also acknowledge Haitao Liu for his support in the Writing – review and editing stage of the paper.

\end{document}